%% file: main.tex
\documentclass[runningheads]{llncs}
\usepackage[T1]{fontenc}
\usepackage{arydshln}
\usepackage{graphicx}
\usepackage{tabularx}
\usepackage{cite}
\usepackage{amsmath,amssymb,amsfonts}
\usepackage{algorithmic}
\usepackage{textcomp}
\usepackage{multirow}
\usepackage{xcolor}
\usepackage{booktabs} 
\usepackage{enumerate}
\usepackage{svg}
\usepackage[colorlinks=true,
            linkcolor=blue,
            anchorcolor=blue,
            citecolor=blue,
            urlcolor=blue]{hyperref}
\usepackage{color}

\usepackage{makecell} 

\begin{document}
\title{Examining GPT's Capability to Generate and Map Course Concepts and Their Relationships}





%
\titlerunning{Generate and Map Course Concepts and Their Relationships based on GPT}
%



\author{
Tianyuan Yang\inst{1}
\and
Baofeng Ren\inst{1}
\and
Chenghao Gu\inst{1}
\and
Tianjia	He\inst{1}
\and
Boxuan Ma\inst{2} 
\and
Shin'ichi Konomi\inst{2}}
\authorrunning{Yang et al.}

\institute{Gurudate School of Information Science and Electrical Engineering, Kyushu University, Fukuoka, Japan\\
\email{\{yang.tianyuan.791, ren.baofeng.817, gu.chenghao.564, he.tianjia.371\}s.kyushu-u.ac.jp}
\and
Faculty of Arts and Science, Kyushu University, Fukuoka, Japan  \\
\email{\{boxuan, konomi\}@artsci.kyushu-u.ac.jp}
}

\maketitle              
\begin{abstract}

Extracting key concepts and their relationships from course information and materials facilitates the provision of visualizations and recommendations for learners who need to select the right courses to take from a large number of courses. However, identifying and extracting themes manually is labor-intensive and time-consuming. Previous machine learning-based methods to extract relevant concepts from courses heavily rely on detailed course materials, which necessitates labor-intensive preparation of course materials. This paper investigates the potential of LLMs such as GPT in automatically generating course concepts and their relations. Specifically, we design a suite of prompts and provide GPT with the course information with different levels of detail, thereby generating high-quality course concepts and identifying their relations. Furthermore, we comprehensively evaluate the quality of the generated concepts and relationships through extensive experiments. Our results demonstrate the viability of LLMs as a tool for supporting educational content selection and delivery.

\end{abstract}
%
%
%


\input{1_introduction}

\input{2_RW}
\input{3_method}
\input{4_evaluation}

\input{5_results}
\section{Conclusion}

We have evaluated GPT’s capabilities in concept generation and relationship identification through extensive experiments, including comparisons with baseline models, ablation studies, and human evaluations. We designed three tasks to comprehensively evaluate GPT’s performance in these areas. The results show that GPT excels in generating high-quality course concepts. Furthermore, GPT can extract low-frequency or generate not appeared concepts that go beyond the input text, addressing gaps in existing methods. In terms of relationship identification, GPT performed strongly, accurately predicting prerequisite relationships that are subtle and often challenging for human evaluators. 

Although we have evaluated course concepts from the perspectives of experts, experts' perspectives can differ from students' perspectives. Thus, our future work includes extensive user studies evaluating the generated concepts in real-world scenarios. Additionally, we plan to explore a wider range of LLMs and investigate their performance across different academic disciplines, particularly for interdisciplinary courses where conceptual and relational boundaries are more fluid.

\bibliographystyle{splncs04}
\bibliography{references}

\end{document}

%% file: 1_introduction.tex
\section{Introduction}

In MOOC environments, learners often have the autonomy to select courses based on their interests and educational objectives. However, the vast array of available courses can make it challenging for students to identify the most suitable courses, to satisfy their diverse needs. To support informed decision-making, it is essential to provide students with the core concepts of a course and the relationships between the concepts. Such information offers valuable insights into the course content and the prerequisites necessary for effective learning \cite{ma2021courseq, ma2024survey}. While educational institutions and MOOC platforms often provide a sharing environment for course materials, syllabi and keywords, faculty members or academic staff have to create these resources manually. This process is both time-consuming and resource-intensive, posing a significant challenge to scalability \cite{pan2017course}.

To address this, researchers have focused on the automatic extraction of course concepts \cite{yu2019course,pan2017course,manrique2018knowledge,lu-etal-2023-distantly} and the relations between these concepts \cite{li2019should,aytekin2024ace,pan2017prerequisite,liang2015measuring} using course information. For example, Lu et al. \cite{lu-etal-2023-distantly} proposed DS-MOCE, which leverages pre-trained language models and discipline-specific embeddings to extract course concepts from MOOCs with minimal manual annotation. Aytekin et al. \cite{aytekin2024ace} presented a machine learning-assisted framework that integrates semantic analysis and expert validation to generate concepts with prerequisite relations. While these approaches effectively identify concepts and their relationships, they exhibit several notable limitations. One major challenge is their heavy reliance on detailed course content \cite{lu-etal-2023-distantly}. Existing studies typically extract course concepts from textual materials and predict relationships based on metrics such as the location, and frequency of concepts within the text, making it difficult to generate high-quality course concepts when limited information is available. Furthermore, these methods focus on explicit textual features rather than conceptual inference, they struggle to generate concepts that may not appear in the text or occur infrequently, even though these concepts are crucial for understanding the course \cite{pan2017course}. Similarly, the identification of inter-conceptual relationships is highly constrained by the availability of conceptual information \cite{sun2024learning}. Previous methods often rely on explicit co-occurrences or external knowledge bases, making them tend to identify only surface-level associations rather than capturing deeper semantic or causal relationships between concepts.


The integration of LLMs, such as GPT, into educational practices has garnered increasing attention as educational paradigms evolve and technology-driven approaches gain prominence. With its ability to generate contextually relevant content and infer implicit relationships, LLMs can overcome the limitations of traditional NLP methods. Previous research has begun to use AI models like GPT to generate course-related information, such as knowledge concepts \cite{gupta2023chatgpt, Yang2024Leveraging,2023eharameasuring}. However, none of these studies have systematically evaluated GPT's ability to generate and extract curriculum concepts, let alone explored its potential for identifying inter-conceptual relationships. In this paper, we explore the feasibility of applying LLMs in the educational domain, with a particular focus on its ability to generate relevant concepts and relations based on course information. By using course data of varying levels of detail as input and designing tailored prompts, we aim to extract course concepts and inter-conceptual relations automatically. The generated concepts are then compared with several baselines, and their quality is evaluated by extensive experiments. Our study addresses the following key research questions:
\begin{itemize}
    \item \textbf{RQ1}: Can GPT generate high-quality course concepts that align with academic standards and effectively capture key course content?
    \item \textbf{RQ2}: How does the level of detail in the input course information impact GPT's quality and accuracy of course concept generation?
    \item \textbf{RQ3}: Can GPT infer and predict prerequisite relationships between concepts?
\end{itemize}

%% file: 2_RW.tex
\section{Related Work}\label{rw}

\subsection{Concept Extraction and Relation Identification}

The concepts and their relationships in learning resources are critical for helping students understand curricula and select appropriate courses. Significant progress has been made in automating the identification of key concepts and relations. 

Foster et al. \cite{foster2012identifying} proposed a semi-supervised learning approach for core concept identification using expert-annotated features, but its reliance on labeled data limited scalability. Manrique et al. \cite{manrique2018knowledge} applied knowledge graphs to rank concepts, yet their approach was constrained by entity-linking quality and knowledge completeness. Pan et al. \cite{pan2017course} introduced an embedding-based graph propagation method for concept extraction, though it struggled with low-frequency concepts and lacked external knowledge integration. Yu et al. \cite{yu2019course} expanded course concepts using external knowledge bases and interactive feedback, but their method suffered from semantic drift and noise. 

The identification of relationships between concepts has been a key area of research, yet it remains a challenging task. Liang et al. \cite{liang2015measuring} proposed Reference Distance (RefD), a link-based metric that utilizes Wikipedia hyperlinks to assess prerequisite relations. Pan et al. \cite{pan2017prerequisite} introduced embedding-based methods to identify relationships in MOOCs, leveraging textual data for relational inference. Manrique et al. \cite{manrique2019exploring} explored the use of general-purpose knowledge graphs, such as DBpedia and Wikidata, to model concept dependencies. Zhang et al. \cite{zhang2022weakly} developed a variational graph autoencoder designed to estimate precedence relations within knowledge graphs. More recently, Aytekin et al. \cite{aytekin2024ace} proposed ACE, a machine learning-assisted approach that integrates expert feedback to construct Educational Knowledge Graphs, significantly reducing the need for manual labeling. Although these methods effectively address concept extraction, they rely heavily on textual data and often incur high computational costs due to complex models. Recent work has integrated large language models (LLMs) and knowledge graphs to enhance concept identification \cite{reales2024core,2023eharameasuring, Yang2024Leveraging}. However, few studies have systematically evaluated GPTs' capabilities in concept generation and relation identification tasks.

\subsection{Large Language Models in Education}

Large Language Models (LLMs), pre-trained on extensive textual data, have become a cornerstone of modern NLP research. Recent advancements have led to high-performing LLMs, such as GPT-3, ChatGPT, and GPT-4, excelling in tasks like machine translation, text summarization, and question-answering \cite{brown2020language}. These models also demonstrate remarkable performance in downstream tasks with minimal or no prompt demonstrations \cite{wei2022emergent,qiao-etal-2023-reasoning,wei2023zero}. Their emergence presents new opportunities in education, including content generation, personalized learning, and tool enhancement\cite{wu2024survey, Yang2024Leveraging}.

Previous research has explored diverse educational applications of LLMs \cite{2024YangMaking,bao2023large,lekan2023ai, Yang2024Boosting}, such as course recommendation, content creation, and addressing data sparsity \cite{wu2024survey, Yang2024Leveraging,2023eharameasuring}. For instance, Barany et al. \cite{barany2024chatgpt} compared manual, automated, and hybrid approaches to qualitative codebook development, and Castleman et al. \cite{castleman2023examining} investigated the integration of knowledge bases into GPT-based tutoring systems. Lin et al. \cite{lin2024can} explored GPT-generated feedback for tutor training, aiming to improve educational tool quality.

While numerous studies have explored the application of GPT's generative capabilities in education, research focusing on using GPT to generate course concepts and identify the inter-conceptual relationships remains scarce and underexplored.  In this work, we leverage ChatGPT to generate course concepts and predict inter-conceptual relationships, exploring its usability in education.

%% file: 3_method.tex
\section{Methodology}\label{3}

\begin{figure}[t]
\centering
\includegraphics[width=\textwidth]{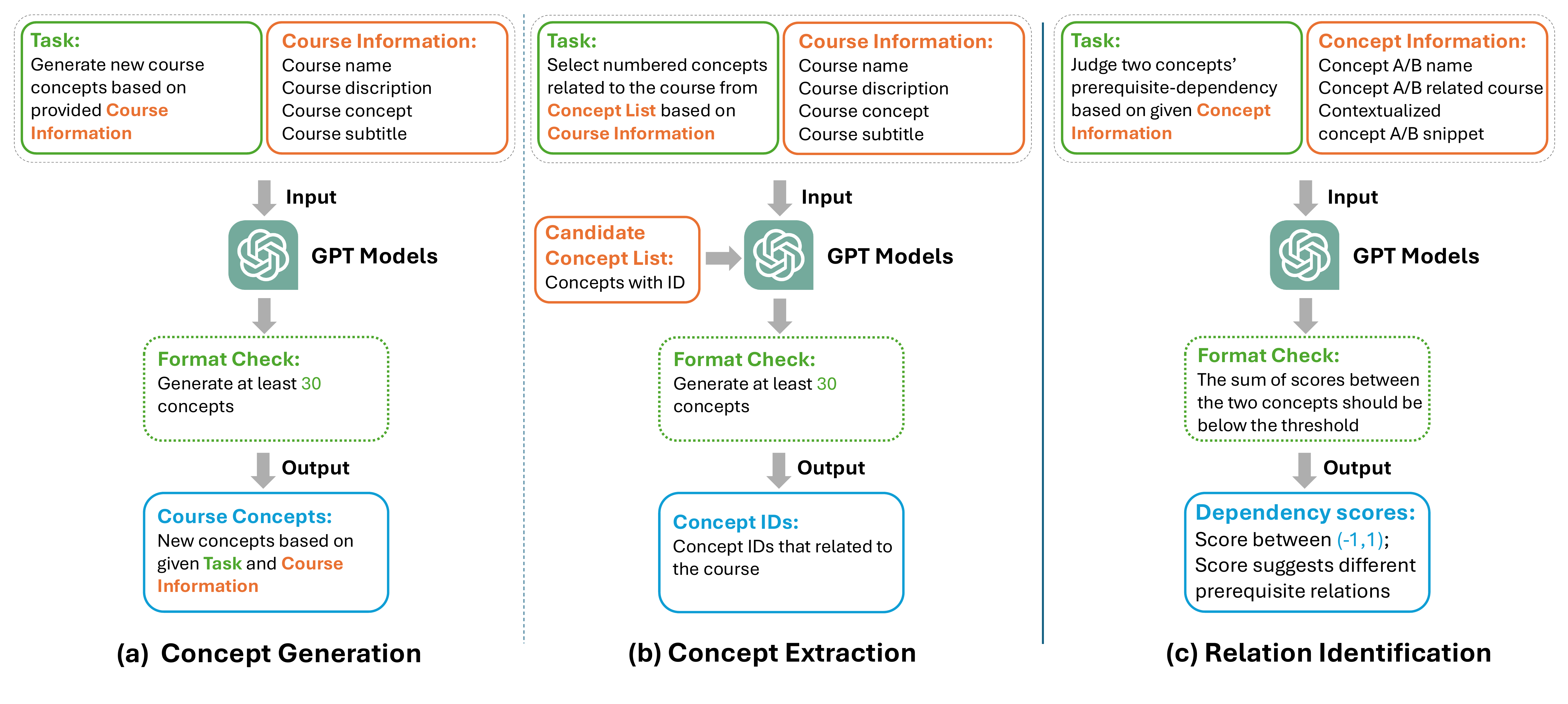}
\caption{Workflow of utilizing GPT to perform two concept-level tasks and one relation-level task. (a) and (b) are concept-level tasks, while (c) is relation-level task.}
\label{workflow}
\end{figure}

The overall workflow for utilizing GPT for concept generation and relation identification is illustrated in Figure \ref{workflow}. Specifically, we developed two concept-level tasks: \textbf{Concept Generation} and \textbf{Concept Extraction}, as well as a relation-level task: \textbf{Relation Identification}. 


\subsection{Concept-level Task Design}
\subsubsection{Concept Generation}

\begin{figure}[t]
\centering
\includegraphics[width=\textwidth]{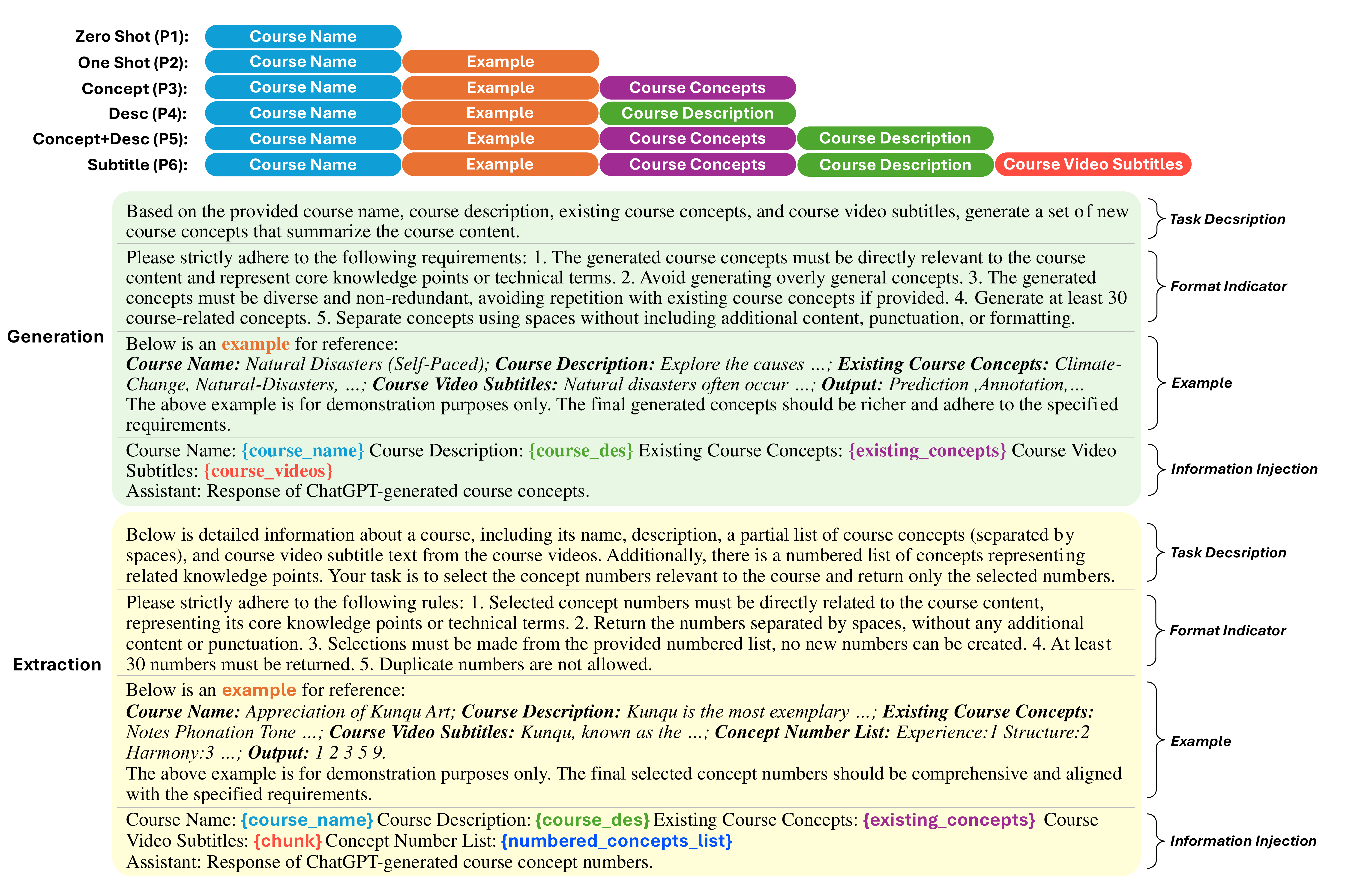}
\caption{Examples of concept generation and extraction task prompts.}
\label{prompt_concept}
\end{figure}

The concept generation task allows GPT to produce outputs based on the input prompt, without strict constraints or predefined answer ranges \cite{reales2024core,lai2024knowledge}. As shown in Fig. \ref{workflow} (a), we utilize GPT to generate relevant concepts for each course. We leverage the target course and its related information and incorporate this information to construct a prompt for GPT. The prompt consists of three components: task description, format indicator, and information injection. As illustrated in Figure \ref{prompt_concept}, the task description specifies the objective, such as concept generation, and outlines the required information. The format indicator defines the desired output structure and the number of concepts to be generated. To ensure GPT meets the specified requirements, a retry mechanism is implemented to verify the output aligns with the prompt's criteria. Information injection provides GPT with relevant course details, including the course name, description, related concepts, video subtitles, and examples.

To systematically assess the impact of varying information granularity on GPT's performance in generating course concepts, we designed six prompts: \textit{Zero-Shot} (P1), \textit{One-Shot} (P2), \textit{Concept} (P3), \textit{Desc} (P4), \textit{Concept+Desc} (P5), and \textit{Subtitle} (P6) to provide GPT with different levels of contextual information. \textit{Zero-Shot} (P1) and \textit{One-Shot} (P2) offer minimal course information, with \textit{Zero-Shot} (P1) providing only the course name and \textit{One-Shot} (P2) including both the course name and an example. \textit{Concept} (P3), \textit{Desc} (P4), and \textit{Concept+Desc} (P5) supply more general course-related details, with \textit{Concept} (P3) providing the course name, an example, and related concepts; \textit{Desc} (P4) offering the course name, an example, and a course description; and \textit{Concept+Desc} (P5) combining the course name, an example, related concepts, and the course description. \textit{Subtitle} (P6) delivers comprehensive course information by extending \textit{Concept+Desc} (P5) to incorporate subtitle information from the course video. By structuring the prompts in this manner, we aim to evaluate how varying levels of information granularity influence GPT's ability to generate relevant course concepts.

\subsubsection{Concept Extraction}

This task requires GPT to select the most appropriate concepts from a set of predefined options provided in the input prompt. The model must comprehend the options, and make accurate, context-based selections. The overall workflow of the proposed method for the concept extraction task is shown in Figure \ref{workflow} (b), and the prompt design can be seen in Figure \ref{prompt_concept}. Similarly, we designed six prompts as in the concept generation task. The difference in this task is that we provide a predefined list of concept candidates, as specified in the prompt, and require GPT to select the concepts from this list that are most suitable for the course. Notably, when designing the prompt, we pre-numbered the concept list and instructed GPT to output the numbers corresponding to the concepts relevant to the course, rather than the concepts themselves to prevent hallucination.

\subsection{Relation-level Task Design}

\begin{figure}[t]
\centering
\includegraphics[width=\textwidth]{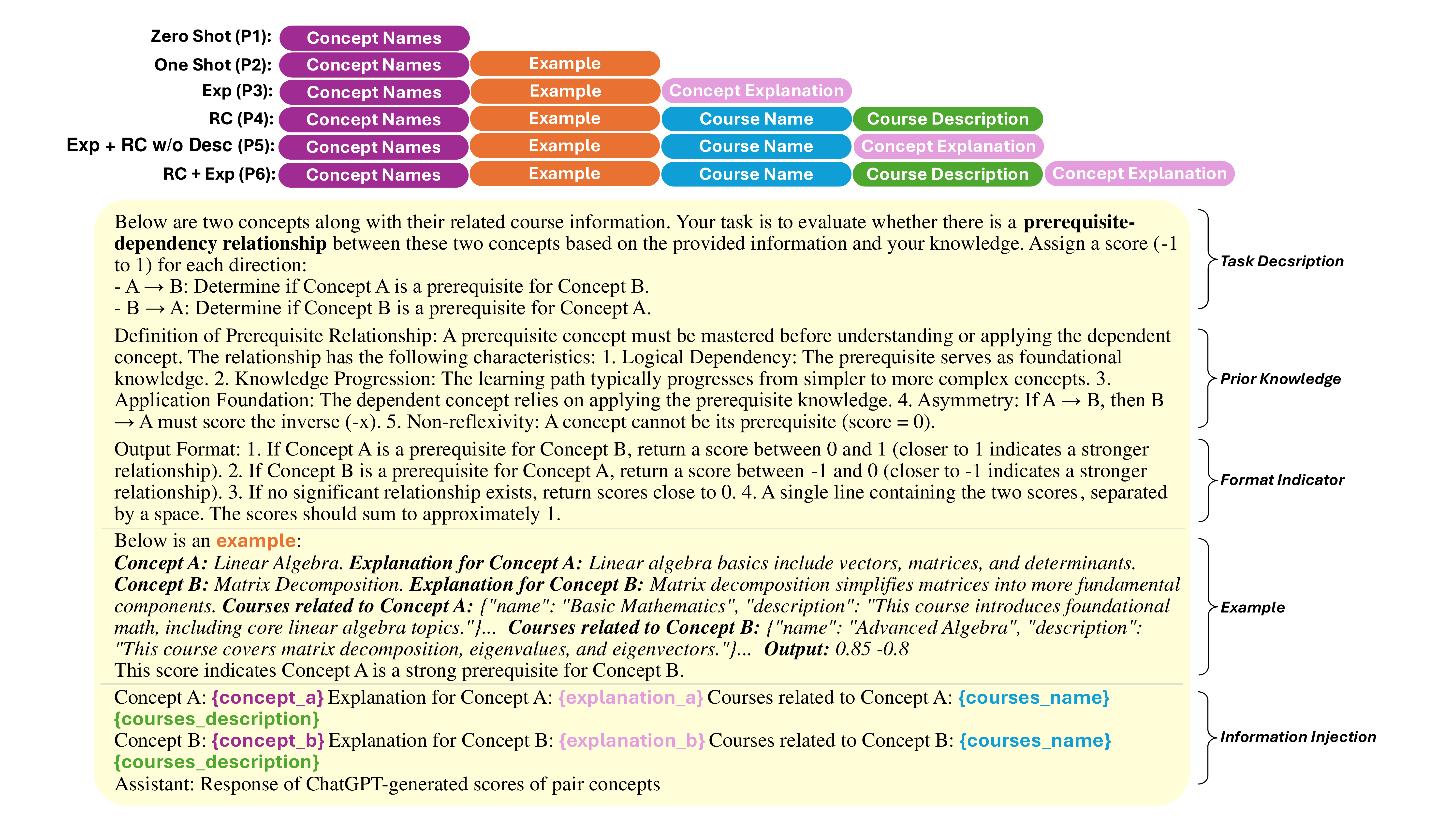}
\caption{Examples of relation identification task prompts.}
\label{prompt_relation}
\end{figure}

Relation identification involves identifying and extracting semantic relations between concepts from curriculum content. Aligned with previous studies \cite{li2019should,aytekin2024ace}, we focus on identifying prerequisite relations. The overall workflow, as illustrated in Figure \ref{workflow} (c), we fed a pair of concepts (\textit{Concept A, Concept B}) into GPT with specifically designed prompts containing varying levels of information about the concepts. GPT outputs a numerical score in the range of -1 to 1, representing the likelihood of a prerequisite relationship between Concept A and Concept B in the pair. The prompt design is illustrated in Figure \ref{prompt_relation}, which comprises four key components: task description, prior knowledge, format indicator, and information injection. The task description instructs GPT to output likelihood scores for two scenarios: Concept A as a prerequisite for Concept B, and vice versa. We designed the prompt based on Liang et al. \cite{liang2015measuring}, which defines prerequisite relationships by stating that if Concept A is a prerequisite for Concept B, the reverse cannot be true. We incorporated this as prior knowledge, ensuring that the likelihood scores of reversed relationships ideally approach zero. To enforce this asymmetry, the format indicator includes a retry mechanism to ensure compliance and specifies that output scores range from -1 to 1, where 1 indicates Concept A is a prerequisite for Concept B, and -1 suggests the opposite. This design effectively mitigates the GPT hallucination issue, ensuring more reliable outputs.

Information injection provides GPT with relevant details about the concept pairs, including their names, contextual snippets, course names, course descriptions, and pertinent examples. Similarly, we designed six prompts with varying levels of information: \textit{Zero-Shot} (P1), \textit{One-Shot} (P2), \textit{Explanation} (P3), \textit{Related Course} (P4), \textit{Explanation+Related Course w/o Desc} (P5), and \textit{Explanation+Related Course} (P6). The \textit{Zero-Shot} prompt provides only the concept pair names, while \textit{One-Shot} adds an example to illustrate the concept relationship. \textit{Explanation} (P3) builds on P2 by incorporating contextualized snippets that explain each concept separately. \textit{Related Course} (P4) includes course names and descriptions associated with the given concepts. \textit{Explanation+Related Course} (P6) integrates both contextualized concept snippets and course information, whereas \textit{Explanation+Related Course w/o Desc} (P5) is similar to P6 but excludes course descriptions. The rationale behind our prompt design is to explore different levels of information granularity. Specifically, P1 and P2 serve as baselines without additional information. P3 introduces contextual explanations of individual concepts, reflecting a concept-level augmentation, while P4 introduces related course information, representing a course-level augmentation.  Finally, P5 and P6 combine both concept-level and course-level information to examine their joint impact on prerequisite prediction.

%% file: 4_evaluation.tex
\section{Experimental Setup}\label{4}

\subsection{Dataset}


We utilized a dataset collected from the XuetangX MOOC platform, provided by Yu et al. \cite{yu-etal-2020-mooccube}. After preprocessing the dataset, it included 683 courses and 25,161 distinct course concepts. Each course in the dataset is associated with a course description and related knowledge concepts. Since each course is presented in video format, the dataset also includes subtitle text corresponding to each course video. Additionally, the dataset includes 1,027 prerequisite relationships between certain concepts. We used these concepts and relationships as the ground truth for evaluating GPT-generated outputs.



\subsection{Baselines}

We tested GPT-3.5, GPT4o-mini, and GPT4o because of their affordability and widespread adoption. We compare the course concepts generated by GPTs and those generated by various NLP methods. We selected three categories of methods for comparison: word frequency-based methods, deep learning-based methods, and graph-based methods. The word frequency-based baselines include PMI \cite{church1990word}, TF-IDF \cite{salton1988term}, and TextRank \cite{mihalcea2004textrank}. The deep learning-based baselines are W2V (Word2Vec) \cite{mikolov2013linguistic} and BERTScore \cite{zhang2019bertscore}, while the graph-based baseline is TPR \cite{liu2010automatic}.

\subsection{Evaluation Metrics}

We used four widely adopted metrics \cite{reales2024core,li2019should}: Precision, Recall, F1 score, and Accuracy to evaluate the performance of GPTs. These metrics are widely used in prior research and were chosen to ensure a comprehensive evaluation. In addition, we employed human evaluation to assess the quality of the generated concepts following previous studies \cite{Yang2024Leveraging,aytekin2024ace}.


%% file: 5_results.tex
\section{Results}\label{5}

\subsection{Performance on Concept Generation}



\begin{table*}[t]
    \centering
    \caption{Performance Comparison$(\%)$ on MOOCCube dataset}
    \begin{tabular*}{\textwidth}{@{\extracolsep{\fill}}llcccc}\toprule
        Category & Method & Precision & Recall & F1 Score & Accuracy \\\midrule
        \multirow{3}{*}{Word Frequency} 
        & PMI       & 2.90 & 0.90 & 1.26 & 0.63 \\
        & TF-IDF    & 16.33 & 2.61 & 4.35 & 2.25 \\
        & TextRank  & 14.78 & 2.07 & 3.60 & 1.85 \\\midrule
        \multirow{2}{*}{Deep Learning}
        & W2V       & 12.17 & 1.56 & 2.75 & 1.43 \\
        & BERTScore & 10.17 & 1.64 & 2.70 & 1.38 \\\midrule
        Graph-based 
        & TPR       & 13.50 & 1.99 & 3.43 & 1.76 \\\midrule
        \multirow{3}{*}{LLMs}
        & GPT-3.5   & \textbf{67.48} & \textbf{39.32} & \textbf{46.38} & \textbf{34.03} \\
        & GPT4o-mini & 13.97 & 19.41 & 15.29 & 9.08 \\
        & GPT4o     & 17.90 & 21.37 & 18.55 & 12.27 \\
        \bottomrule
    \end{tabular*}
    \label{comparison}
\end{table*}

\textbf{Performance Comparison} We compared our approach with highly influential methods in the NLP domain by providing the subtitled text of the course video to each baseline, which aligns with our prompt \textit{Subtitle} (P6). We selected 100 courses and generated at least 30 concepts for each course using both GPT and NLP-based methods. The results are shown in Table \ref{comparison}. We can see that GPTs demonstrate exceptional performance, largely due to their superior reasoning ability compared to existing NLP techniques. NLP methods generate concepts strictly based on the text, meaning that if a concept is infrequent or absent from the subtitles, these approaches fail to produce it. In contrast, GPT-generated concepts go beyond the textual content itself, leveraging broader contextual understanding and reasoning. For instance, in a machine learning course, traditional models primarily extract terms explicitly mentioned in the subtitles, such as ``\textit{gradient descent}” or ``\textit{neural networks}”. However, GPT is capable of generating higher-level concepts like ``\textit{bias-variance tradeoff}” or ``\textit{Bayesian inference}”, even if they are not explicitly stated.

\begin{figure}[t]
\centering
\includegraphics[width=\textwidth]{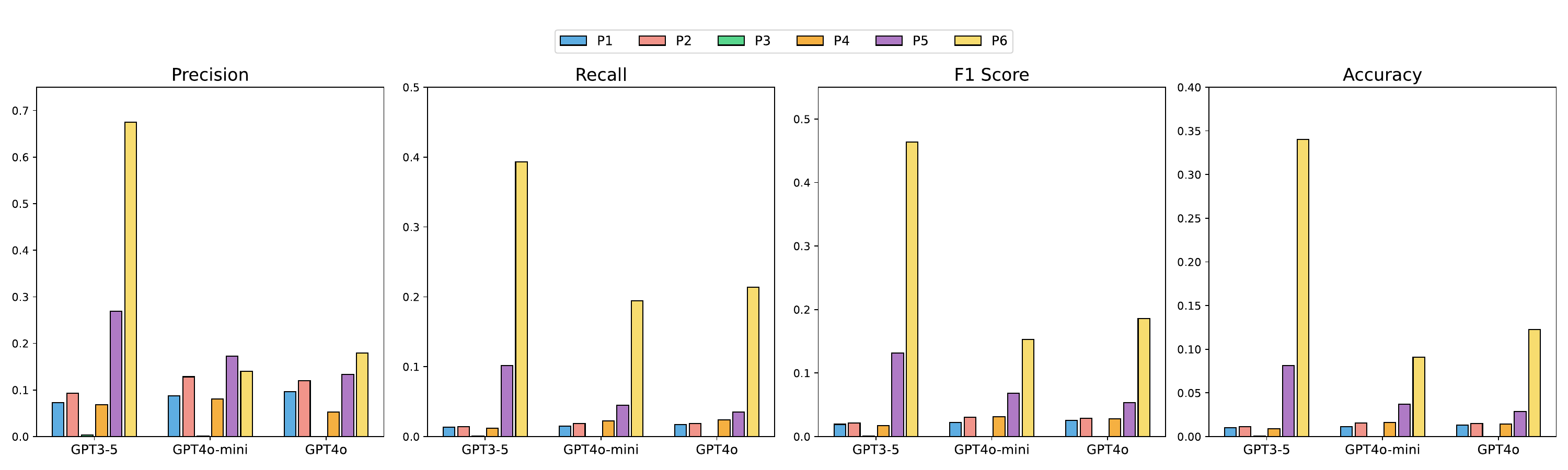}
\caption{Performance comparison of different GPT models and prompts.}
\label{open_ended}
\end{figure}

\textbf{Ablation Study on Concept Generation} We then evaluated the performance of three different GPT models across six prompts, comparing the generated concepts with the Ground Truth in the dataset. The results are shown in Figure \ref{open_ended}, and our analysis yielded several observations: (1) As the amount of information increases (P5, P6), the performance of all three GPT models improves, indicating that additional information enhances GPT’s ability to generate concepts. (2) Contrary to expectations, GPT-3.5 achieved the best results. After looking into the Ground Truth of the dataset, and examining the concepts generated by GPT, we found that the concepts generated by GPTs were not fully consistent with the Ground Truth, while they do not appear to be low quality. This is also aligned with previous work \cite{liu2023chatgpt, Yang2024Leveraging} which suggests that GPTs can generate high-quality content that is informative and meaningful, even if it does not strictly adhere to predefined ground truth labels. Therefore, the evaluation is insufficient and we also conducted a human evaluation to assess GPT’s concept generation capabilities following previous work \cite{liu2023chatgpt, Yang2024Leveraging}.

\textbf{Human Evaluation on Concept Generation} We enlisted four experts familiar with these courses to evaluate the quality of the generated course concepts using a 5-point scale based on their relevance to the course and accuracy. We tested three models: GPT-3.5-turbo, GPT-4o-mini, and GPT-4o. Each generated concepts for 100 courses using various prompts. From each model-prompt combination, we sampled 20 courses and randomly selected 10 concepts per course. Experts were asked to evaluate both the GPT-generated concepts and a set of ground truth concepts to compare their quality.

\begin{table*}[t]
    \centering
    \caption{Average Scores of Human Evaluation for Each Prompt and Model.}
    \begin{tabularx}{0.95\textwidth}{lXXXXXXXX}\toprule
     
        Model/Prompt & P1 & P2  & P3 & P4 & P5 & P6  \\\midrule
        Ground Truth & 2.677 & 2.677 & 2.677 & 2.677 & 2.677 & 2.677 \\ \midrule
        GPT-3.5  & 3.613 & 3.454 & \textbf{3.630} & 3.346 & 3.083 & 3.205 \\ 
        GPT4o-mini & 3.620 & 3.461 & 3.320 & 3.376 & 3.341 & 3.276  \\ 
        GPT4o & \textbf{3.700} & \textbf{3.516} & 3.478 & \textbf{3.435} & \textbf{3.519} & \textbf{3.573} \\
        \bottomrule
    \end{tabularx}
    \label{expert}
\end{table*}


\begin{figure}[t]
\centering
\includegraphics[width=\textwidth]{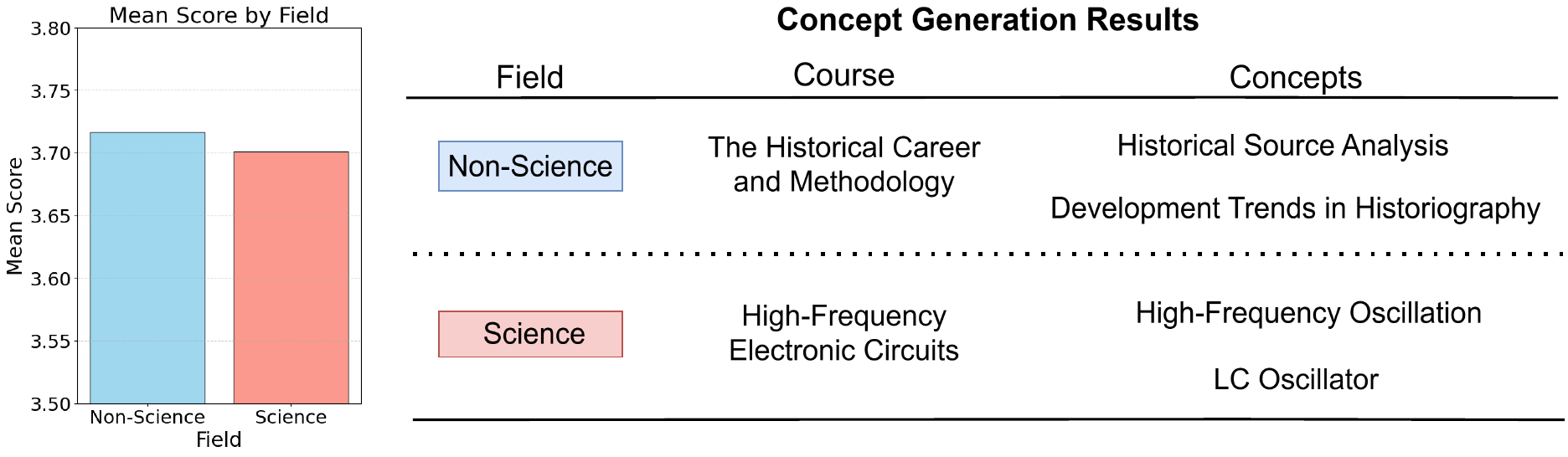}
\caption{Human evaluation of the science and non-science courses, along with examples of concept generation in both fields.}
\label{field}
\end{figure}

The mean scores of human evaluations, shown in Table \ref{expert}, reveal several insights: (1) ChatGPT-generated concepts surpass the quality of Ground Truth included in the MOOCCube dataset, which is extracted by neural networks and manually. (2) As the GPT model improves, the quality of generated concepts increases. ChatGPT-4o, the most advanced model, consistently generates high-quality concepts. (3) When detailed information is provided, the efficiency of processing varies significantly across models. Interestingly, simpler prompts with less information yield better results for the lower-tier models, as excessive details overwhelm their processing capabilities. Moreover, the large volume of course video subtitles often contains excessive noise. (4) Contrary to expectations, \textit{Zero-Shot} prompts receive higher scores from experts. While providing more detailed input might intuitively seem beneficial, in \textit{Zero-Shot} and \textit{One-Shot} scenarios, GPT tends to generate general and coarse-grained concepts due to limited course-specific information.

We also conducted brief interviews with four experts to gather their insights. First, all four experts expressed a shared preference for the granularity of concepts, agreeing that overly fine-grained concepts might be difficult for students to understand, particularly those new to the subject. Second, the experts unanimously agreed that the quality of GPT-generated concepts exceeded their expectations. Finally, Expert A observed a distinct contrast in the granularity of concepts generated by GPT across disciplines. For science courses, the generated concepts were highly detailed and technical, while for no-science courses, the concepts were broader and more generalized. As shown in the left part of Figure \ref{field}, we compared the average expert scores of non-science courses and science courses, finding that non-science courses received slightly higher scores. On the right side of Figure \ref{field}, a non-science course such as \textit{The Historical Career and Methodology} produced concepts like \textit{Development Trends in Historiography} and \textit{Historical Source Analysis}, which are comprehensive and overarching. In contrast, a science course like \textit{High-Frequency Electronic Circuits} yielded concepts such as \textit{High-Frequency Oscillation} and \textit{LC Oscillator}, which delve into specific technical details.




\subsection{Performance on Concept Extraction}

In addition to the generation task, we designed the Concept Extraction Task to evaluate GPT's reasoning ability. In this task, we predefined a list of candidate concepts and asked GPT to select the concepts relevant to the target course. Specifically, the candidate concept list was constructed as the union of all concepts from the target course and another course. We employed two strategies for constructing the candidate concept list to test GPT's ability to distinguish concepts. In the first strategy, we randomly selected a course from a different field than the target course, where most concepts had clear distinctions from the target course. In the second strategy, we randomly sampled a course within the same field as the target course, resulting in a list that included concepts more similar to those of the target course. Same to the concept generation task, we tested three GPT models and six prompt variations for this task.

\begin{figure}[t]
\centering
\includegraphics[width=\textwidth]{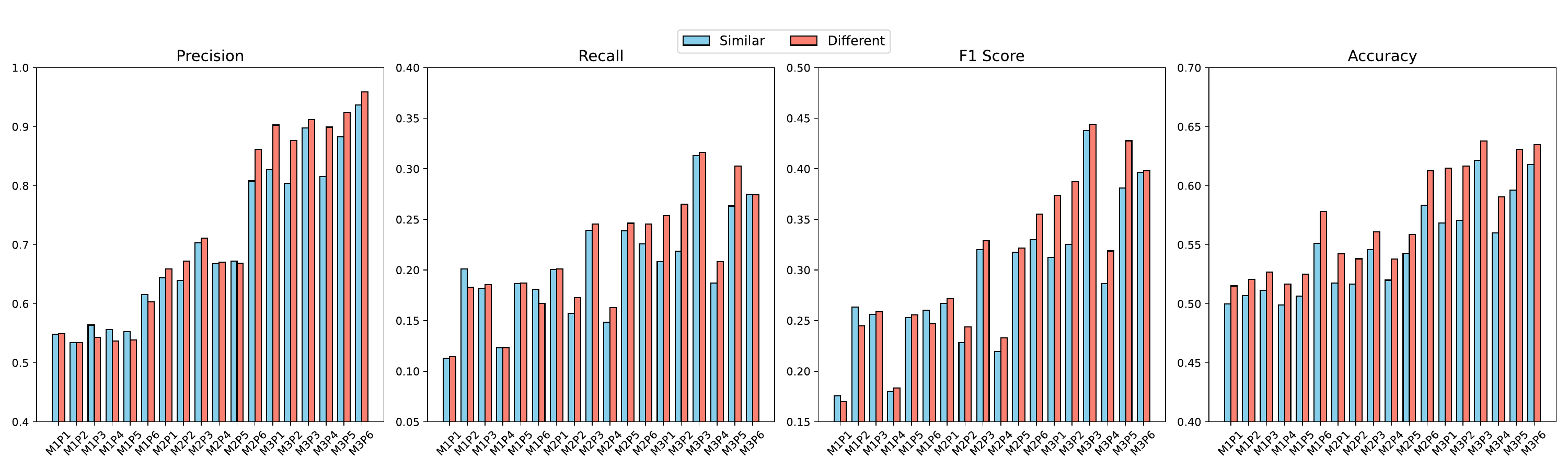}
\caption{Performance comparison of different GPT models and prompt configurations on concept extraction task. M1, M2, and M3 denote GPT3.5, GPT4o-mini, and GPT4o, respectively.}
\label{multi_choice}
\end{figure}

\begin{figure}[t]
\centering
\includegraphics[width=\textwidth]{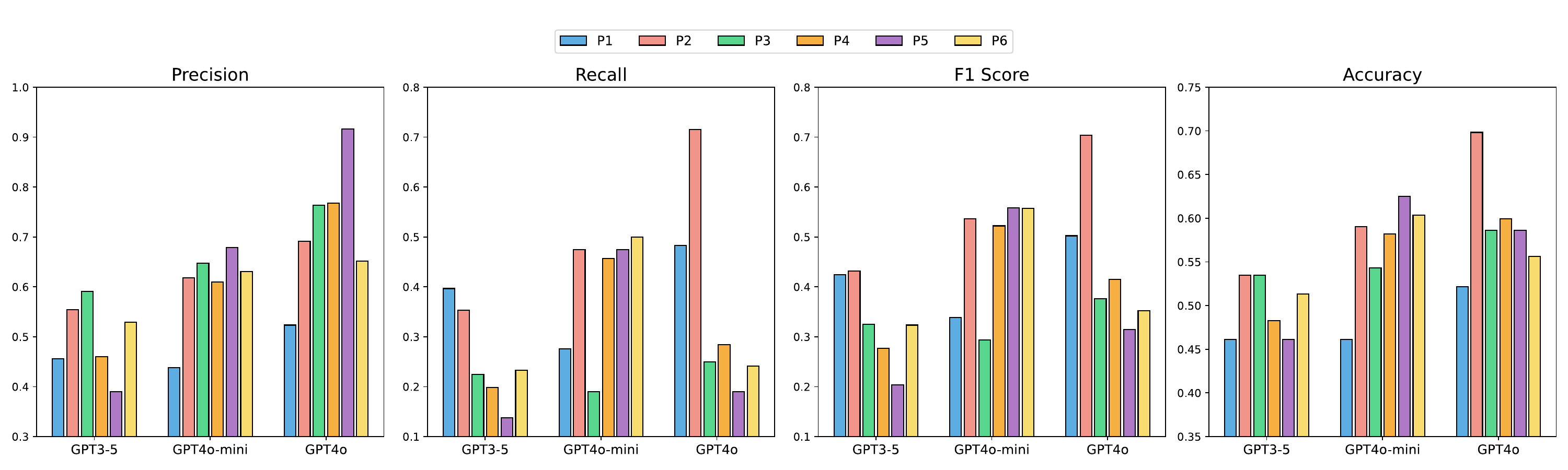}
\caption{Performance comparison of different GPT models and prompt configurations on inter-conceptual relation identification task.}
\label{relation}
\end{figure}

The results presented in Figure \ref{multi_choice}, lead to several key observations: (1) Among the three GPT models, GPT-4o demonstrates the strongest reasoning ability, achieving the highest performance across all four evaluation metrics, indicating its effectiveness in extracting course concepts. (2) Increasing input information generally improves performance, with P3, P4, P5, and P6 outperforming P1 and P2. However, in GPT-3.5 and GPT-4o-mini, excessive information can introduce noise that hinders model inference, leading to worse performance. (3) GPTs perform better when the candidate concept list consists of concepts from the target course and a course from a different domain, aligning with expectations since the distinction between concepts is clearer. However, GPTs also perform well when selecting from a candidate list composed of concepts from the target course and another course within the same domain, demonstrating their ability to process similar information effectively.



\subsection{Performance on Relation Identification}


In addition to the concept-level task, we utilized GPT to identify relationships between concepts. In this experiment, we tested three models, GPT-3.5-turbo, GPT4o-mini, and GPT4o. Each GPT model was applied to 100 concept pairs using different prompts to determine the existence of prerequisite relationships. The results, presented in Figure \ref{relation}, lead to several observations: (1) Performance improves as the model advances, with GPT-4o achieving the best overall performance. (2) Similarly, the performance of GPT tends to improve with an increase in input information. However, increasing the amount of information does not always lead to performance gains across all models. This outcome can be attributed to two factors: first, the prerequisite relationships between many concept pairs are inherently subtle and challenging to identify, even for human evaluators. Second, the provided input information may not always include the critical details necessary to determine the existence of a prerequisite relationship, limiting the model's ability to make accurate judgments.